\documentclass[twocolumn]{aastex6}
\usepackage{amsmath}
\usepackage{color}
\usepackage{multirow}
\newcommand{\rom}[1]{\textup{\uppercase\expandafter{\romannumeral#1}}}
\newcommand{\mcr}[1]{\multicolumn{1}{r}{#1}}
\newcommand{\nd}{\multicolumn{1}{c}{$\dots$}}
\usepackage[normalem]{ulem}
\shorttitle{Miras in M33}
\shortauthors{Yuan, He, Macri, Long \& Huang}

\begin{document}
\title{The M33 Synoptic Stellar Survey. \rom{2}. Mira Variables}

\author{Wenlong Yuan\altaffilmark{1}, Shiyuan He\altaffilmark{2},
  Lucas M.~Macri\altaffilmark{1,3}, James Long\altaffilmark{2}, \& Jianhua Z.~Huang\altaffilmark{2}}
\altaffiltext{1}{George P.~and Cynthia W.~Mitchell Institute for Fundamental Physics \& Astronomy,\\Department of Physics \& Astronomy, Texas A\&M University, College Station, TX, USA}
\altaffiltext{2}{Department of Statistics, Texas A\&M University, College Station, TX, USA}
\altaffiltext{3}{Corresponding author, {\tt lmacri@tamu.edu}}

\begin{abstract} 
We present the discovery of 1847 Mira candidates in the Local Group galaxy M33 using a novel semi-parametric periodogram technique coupled with a Random Forest classifier. The algorithms were applied to $\sim 2.4\times 10^5$ $I$-band light curves previously obtained by the M33 Synoptic Stellar Survey. We derive preliminary Period-Luminosity relations at optical, near- \& mid-infrared wavelengths and compare them to the corresponding relations in the Large Magellanic Cloud.
\end{abstract}


\section{Introduction}
Mira variables (Miras) are asymptotic giant branch (AGB) pulsating stars that exhibit large cyclical variations in flux at optical wavelengths, typically with periods spanning $100-700$~d but in extreme cases going beyond 1500~d. The ``canonical'' empirical classification requires $\Delta V\!>\!2.5$~mag within a pulsation period and spectroscopic confirmation \citep{Kholopov1985}. Recent surveys for these variables \citep[such as][]{Soszynski2009} have adopted $\Delta I\!>\!0.$8~mag as the only requirement for classification, since spectroscopic followup of very large samples is not currently feasible. Longer-term variations in the mean flux level of each cycle are typical \citep{Mattei1997,Whitelock1997} and visual light curves exhibit a wide range of shapes; \citet{Ludendorff1928} classified Miras into three classes and ten subclasses based on this attribute.

Since the progenitors of Miras are relatively low-mass stars, they are ubiquitous and present in all types of galaxies. Thousands of Mira candidates have been discovered in the Milky Way and the Magellanic Clouds based on photometry from the Optical Gravitational Lensing Experiment \citep[hereafter, OGLE;][]{Udalski1992} and MACHO \citep{Alcock1993} projects. The Mira Period-Luminosity relations (hereafter, PLRs) were initially studied by \citet{Gerasimovic1928}. \citet{Glass1981} found the first evidence of a near-infrared (NIR) PLR for Miras, based on a small sample of variables in the Large Magellanic Cloud (LMC). Using the MACHO database, \citet{Wood1999} were the first to identify multiple PLRs for AGB stars and to confirm the nature of Miras as radial fundamental mode pulsators. Also using MACHO periods, \citet{Glass2003} determined that the Mira $K$-band PLR exhibits a relatively small scatter of $\sigma\sim$0.13~mag while \citet{Whitelock2008} found similar dispersions for $K$-band PLRs separated into O- and C-rich subtypes ($\sigma=0.14$ and 0.15~mag, respectively). These values are comparable to the intrinsic dispersion of the Cepheid PLR in the same bandpass \citep[$\sigma=0.09$~mag,]{Macri2015}. \citet{Soszynski2009, Soszynski2011, Soszynski2013} characterized the NIR Mira PLRs in the LMC, the Small Magellanic Cloud and the Galactic Bulge, respectively, using OGLE and 2MASS photometry. In just a few years, the Large Synoptic Survey Telescope (LSST) will begin to obtain frequent images of dozens of nearby galaxies, which will have the necessary depth to enable the discovery of Miras and the determination of distances to these systems.

In this work we report the results of a search for Mira candidates in M33 using $I$-band observations spanning nearly a decade, obtained by the DIRECT project \citep[][hereafter, M01]{Macri2001} and by \citet[][hereafter, PM11]{Pellerin2011}. Traditional periodogram methods such as Lomb-Scargle \citep{Lomb1976,Scargle1982} are not optimal for this search due to relatively sparse temporal sampling, large gaps between observing seasons, and the expected long-term variations in Mira light curves. We developed a novel semi-parametric periodogram technique \citep[][hereafter, H16]{He2016} based on the Gaussian Process method that contains a data-driven component in the model light curve to account for deviations from strict periodicity and gives an overall better performance. We coupled this algorithm to Random Forest classifiers, training and testing them extensively on simulated light curves.

The rest of the paper is organized as follows: \S\ref{sec:observations} introduces the observations and data reduction; \S\ref{sec:simulation} gives details of the light curve simulations; \S\ref{sec:model} discusses the methodology used to search for Mira candidates and estimate their periods; \S\ref{sec:results} presents our results, which include a comparison of the Random Forest classification with other techniques and preliminary PLRs for O-rich Mira candidates.

\section{Observations and data reduction}\label{sec:observations}

\begin{figure}
\includegraphics[width=0.49\textwidth]{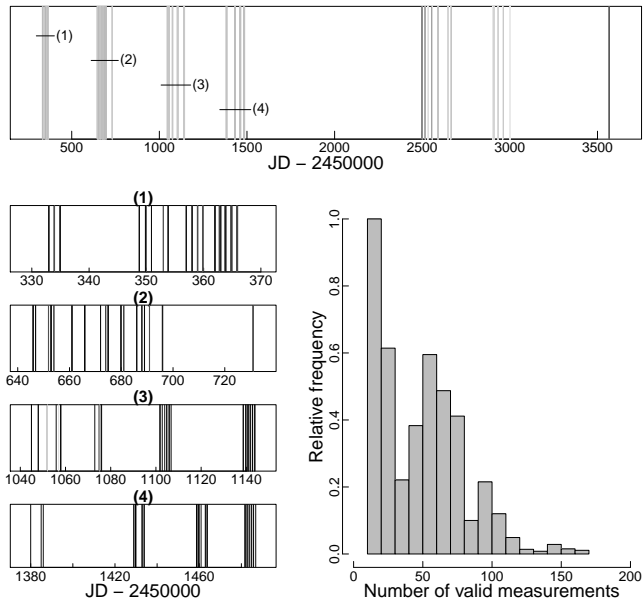}
\caption{{\it Top:} Cadence of M33 observations in $I$ by M01 \& PM11. The grayscale levels are linearly proportional to the number of measurements per sq.~arcmin.~of each epoch. {\it Bottom-left:} Expanded view of the cadence for seasons $1-4$. {\it Bottom-right}: Histogram of measurements for stars with $N\!>\!10$ and $I\!<\!21.45$~mag.\label{fig:obs}}
\vspace*{-18pt}
\end{figure}

We based our search on the observations of M33 obtained by \citet{Macri2001} and \citet{Pellerin2011}. These surveys covered most of the disk of this galaxy with a combined baseline of nearly a decade (1996 September to 1999 November for M01; 2002 August to 2006 August for PM11) mainly using the Fred L. Whipple Observatory 1.2-m and the WIYN 3.5-m telescopes with a variety of cameras (see the respective publications for details). While images were obtained in multiple bandpasses ($BVI$), our analysis is only based on the $I$-band time-series photometry because Mira candidates fall below the detection limit in the bluer bands. Given that both studies had to rely on multiple pointings to cover the area of interest and not all locations were observed on a given night, the sampling pattern varies considerably across the disk. Fig.~\ref{fig:obs} shows the overall sequence of observations, of which only a subset will be applicable at a given position.

We performed new photometric measurements on the pre-processed images from PM11 to mitigate issues arising from geometric distortions and poor image registration at the corners of each field (which corresponds to a single telescope and camera). Unlike the approach of the previous work, we first analyzed the images of a given field and later combined the photometry for matching sources. We obtained aperture and point-spread function (PSF) photometry using the {\tt DAOPHOT}, {\tt ALLSTAR} and {\tt ALLFRAME} programs \citep{Stetson1987,Stetson1994} in a quasi-automatic manner by integrating the tasks into an {\tt R} script pipeline. We defined the PSF for each image using the top 50 bright and isolated stars and selected the one with the sharpest PSF from each field to serve as reference for {\tt ALLFRAME}. We selected a larger number of secondary standards for image registration and to tie the photometric measurements. These were among the brightest few percent of all sources in a given field and had photometric uncertainties below 0.02~mag. We determined frame-to-frame zeropoint offsets, computed mean instrumental magnitudes and extracted light curves using {\tt TRIAL} \citep{Stetson1996}.

We obtained astrometric and photometric calibrations for each field using the catalog published by the Local Group Galaxies Survey \citep[hereafter LGGS,][]{Massey2006}. We derived the astrometric solution for the reference frame of each field using WCSTools \citep{Mink1999}. We matched LGGS sources to the star list from the (now astrometrically calibrated) reference frame of each field and solved the following transformation equation:
\begin{equation}
I_c = (1+a)\cdot I_i + b
\end{equation}
\noindent{where $I_c$ is the magnitude in the standard system (Kron-Cousins $I$ for LGGS) and $I_i$ is the instrumental PSF magnitude of the reference frame of a given field, $b$ is the zeropoint offset and $a$ provides a simple correction for color terms and/or photometric biases due to crowding (given the considerable variation in stellar density across the disk and in image quality among the fields). We were not able to apply a traditional photometric transformation with zeropoint and color terms because we only have single-band ($I$) photometry for the vast majority of the sources. We solved for the coefficients using the top 25\% and 10\% brightest stars in fields imaged at WIYN and FLWO, respectively, applying an iterative outlier rejection of 3 \& 2.3$\sigma$, respectively. The median value of $a$ was 0.001, with 95\% of the values falling between $-0.024$ and $+0.015$.}

\begin{figure}
\vspace*{-12pt}
\includegraphics[width=0.49\textwidth]{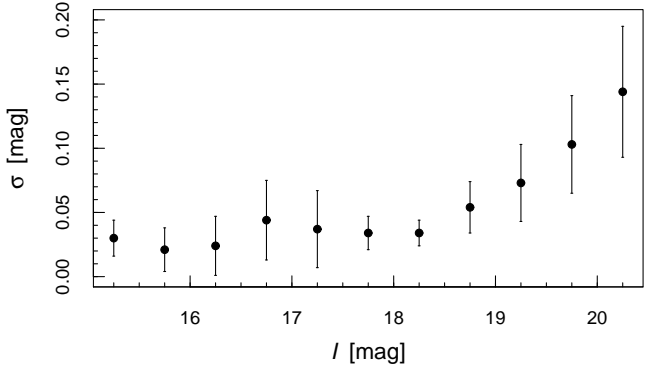}
\caption{Photometric precision for secondary standards as a function of magnitude.\label{fig:photprec}}
\vspace*{-12pt}
\end{figure}

Given the significant overlap between the fields of M01 and PM11, most objects have multiple light curve segments that were merged as follows. If two sources in different fields had coordinates that matched to better than $1\arcsec$ and there were no other sources detected within $1\arcsec$, they they were considered as the same object. If there were neighboring sources within that radius, then the closest object with a magnitude difference less than 0.5~mag was selected (recall that each field was already transformed to the standard system prior to this step). We ensured that at most one source in one field could merge with one source in another field. We tested the photometric precision of the aforementioned steps by reidentifying the local standards of each field and examining the dispersion of the merged light curves relative to the dispersion of individual segments. Fig.~\ref{fig:photprec} shows that we reach a photometric precision of 0.03~mag.

We selected 239907 light curves for the Mira search, rejecting any with less than 10 measurements or with mean magnitudes fainter than $I\!=\!21.45$~mag. The first cut is based on extensive testing via simulated light curves (\S\ref{sec:simulation}) of our algorithm (\S\ref{sec:model}); the procedure does not yield reliable periods for sparser samplings. The second cut is due to the large photometric uncertainties beyond that magnitude limit, which prevent the detection of light curve variations of the expected amplitude.

\section{Simulated M33 light curves} \label{sec:simulation}

We simulated $10^5$ light curves of Miras and the same number of semi-regular variables (SRVs) that accurately reproduce the photometric uncertainties and temporal sampling of the M33 dataset. The simulated light curves were used to test our period determination algorithm and to train a classifier to identify Mira candidates. The methods used in the simulation are based on the ones we developed for H16 and rely on very high precision $I$-band light curves sampled at hundreds of epochs over 7.5 years by phase \rom{3} of the OGLE project \citep{Udalski2008}. We also generated an equal number of artificial light curves of ``constant'' stars, in order to properly balance the training data for the classifier.

\subsection{Miras}\label{ssec:simmira}

The procedure used to fit templates to the OGLE-\rom{3} Mira light curves is explained in detail in \S4.3 of H16. Briefly, the light curve is decomposed into a mean value, a regular variation of period $P$, a low-frequency (long-term) trend and a high-frequency/small-scale term. The latter three components are modeled by a Gaussian process with different kernels. The maximum likelihood method is used to obtain the model parameters as a function of the trial value of $P$. Once the best-fit model is found, it can be used to predict the magnitude at any time $t$ during the observation baseline, including the brightest value reached by the variable (hereafter, $I_m$). This quantity is of interest because \citep{Kanbur1997} has suggested that PLRs at maximum light may have a significantly smaller dispersion than at mean light.

Simulated light curves were generated by sampling the best-fit model using randomly-selected observing patterns from the actual light curves, with equal probabilities. We shifted the starting point of each simulated light curve by a random value $\Delta t$, limited in range only to ensure the resulting light curve was still contained within the span of the OGLE observations. These random shifts helped to obtain many unique simulated light curves when using the same template. We applied a shift of $\Delta m\!=\!+6.2$~mag to account for the approximate difference in apparent $I$-band distance moduli between the LMC and M33 \citep[$\Delta\mu_0\!=\!6.26\pm0.03$~mag and $\Delta A_I=-0.05$~mag based on][]{Pellerin2011,Schlafly2011}. Furthermore, we introduced a realistic amount of photometric noise following the procedure outlined in \S4.1 of H16.

As a final step in our simulations, we took into account the fact that the OGLE LMC observations reach substantially deeper in terms of absolute magnitude than the M33 observations and considered the possibility that the light curve shape of Miras may be a function of their luminosity. If the latter is true, a mismatch of the luminosity functions would bias our classifier. We derived the completeness function of the M33 photometry (by fitting the observed luminosity function with an exponential) and applied it to the luminosity function of the OGLE LMC Mira candidates, after offsetting the latter by the difference in apparent distance moduli between the two galaxies. We then randomly selected simulated Mira light curves such that we reproduced the observed luminosity function of the M33 photometry.

\begin{figure}[t]
\includegraphics[width=0.49\textwidth]{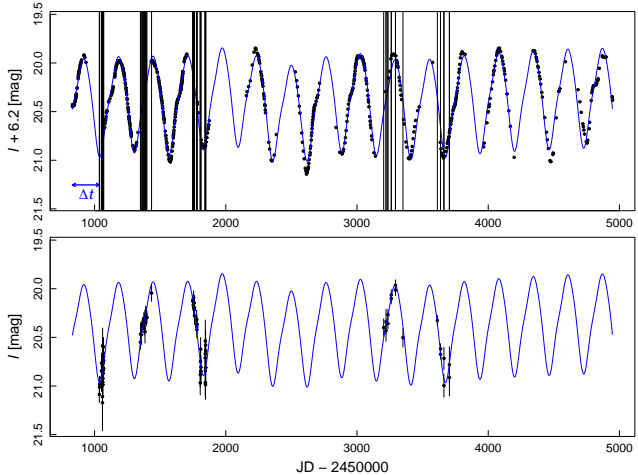}
\caption{Example of a template Mira light curve and simulated M33 measurements. {\it Top:} OGLE measurements of a Mira candidate in the LMC (black points), best-fit template using our model (blue curve), and sampling pattern of one of the M33 fields (vertical black lines). The horizontal blue arrow indicates the random shift applied to the pattern to sample the light curve. {\it Bottom:} Corresponding simulated M33 light curve, including additional photometric noise.\label{fig:sim_mira}}
\end{figure}

\begin{figure}[t]
\includegraphics[width=0.49\textwidth]{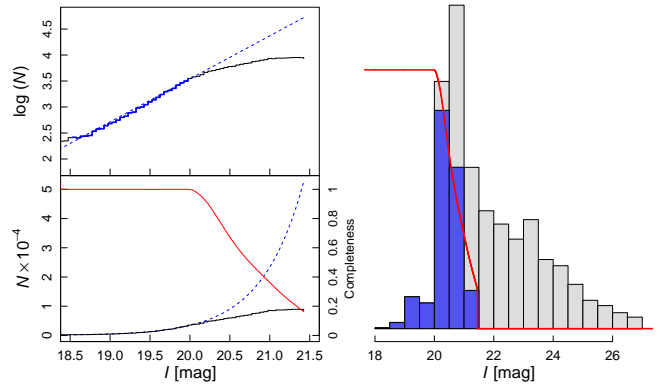}
\caption{{\it Left}: Derivation of the empirical completeness function for M33 photometry (top: logarithmic; bottom: linear scale). An exponential model is fit to the observed luminosity function (solid black line) over the magnitude range shown with a solid blue line and extrapolated over the range plotted with a dotted blue line. The derived completeness function (solid red line) is shown in the bottom panel only. {\it Right}: Magnitude distribution of Mira template light curves before (grey) and after (blue) convolution with the completeness function (red line). \label{fig:complete}}
\end{figure}

Fig.~\ref{fig:sim_mira} shows a representative example of a simulated Mira light curve that mimics the cadence and photometric precision of the actual M33 data, while Fig.~\ref{fig:complete} shows the completeness function of the M33 photometry.

\subsection{Semi-regular variables and ``constant'' stars}

The light curves of SRVs share some similarities with those of Miras (cyclic variations), although they tend to be more chaotic, less periodic, and usually exhibit smaller amplitudes. Nevertheless, since they outnumber Miras 6 to 1 in the catalog of \citet{Soszynski2009}, ignoring them could significantly bias our classifier. Hence, we included simulated SRV light curves in the training data. We obtained templates by applying a smoothing spline to the OGLE observations and generated artificial light curves by following the same procedure as for Miras (sampling based on the M33 observing patterns, random shifts of the starting point, convolution with M33 completeness function, and addition of photometric noise).

Lastly, in order to balance the various types of objects that are used to train the classifier, we simulated light curves of ``constant'' stars by randomly shuffling the observation times of all light curves in our dataset while keeping the original magnitudes and uncertainties. The shuffling removes any potential periodicity in the original data and allows the generation of multiple artificial light curves from the same object.

\section{Semi-parametric model for identification and period determination of Miras} \label{sec:model}

Given the stochastic variations exhibited by Mira light curves at optical wavelengths, traditional algorithms become less efficient at discovering these objects and obtaining reliable periods in the limit of sparsely-sampled observations. We have shown in H16 that our semi-parametric model gives an overall improvement for period recovery in this regime. We applied this model to the M33 observations and coupled it to a Random Forest classifier to identify Mira candidates.

We refer interested readers to \S3 of H16 for a detailed description of our semi-parametric model and its performance, which are only briefly summarized here. The model is based on Gaussian Process regression, which

\noindent{has been previously applied to astronomical time-series observations. For example, \citet{Faraway2016} modeled the light curves of several types of transient event from the the Catalina Real-time Transient Survey with a squared exponential kernel, while \citet{Aigrain2016} applied this technique to {\it Kepler} data to correct systematic trends in its photometry.}

The semi-parametric model we used is a simplified version of the one described in \S\ref{ssec:simmira} to account for the quality of the M33 data. Given a set of measurements over $n$ epochs, $\{(t_i, m_i, \sigma_i)\}_{i=1}^n$, with $t$, $m$, and $\sigma$ representing time, magnitude and measurement uncertainty, respectively, the model is
\begin{equation}
m_i = m + \beta_1\cos(2\pi ft_i) + \beta_2\sin(2\pi ft_i) + h(t_i) + \sigma_i\epsilon_i
\end{equation}
where $f$ is the frequency (d$^{-1}$), $h(t)$ is modeled by a Gaussian Process with the squared exponential kernel,
\begin{equation}
k(t_1,t_2) = \theta_1^2\exp (-{(t_1-t_2)^2}/{2\theta_2^2}),\nonumber
\end{equation}

\noindent{and the amplitude of the periodic component is $A_P=2(\beta_1^2+\beta_2^2)^{1/2}$.\ \ The\ \ parameters $m$,\ \ $\beta_1$, and $\beta_2$ are assumed to follow Gaussian priors and integrated out of the likelihood function when estimating other parameters. Optimization is performed over hyper-parameters  $\theta_1$ and $\theta_2$ for each trial value of $f$, ranging from $5\times 10^{-4}$ to $10^{-2}$ every $10^{-5}$. The log-likelihood function $Q$ (hereafter, ``frequency spectrum'') is evaluated at each trial frequency (see Equation~10 in H16).}

We applied the model to all simulated and real light curves and\ \  adopted the highest\ \ peak in the frequency

\begin{figure*}[t]
\includegraphics[width=\textwidth]{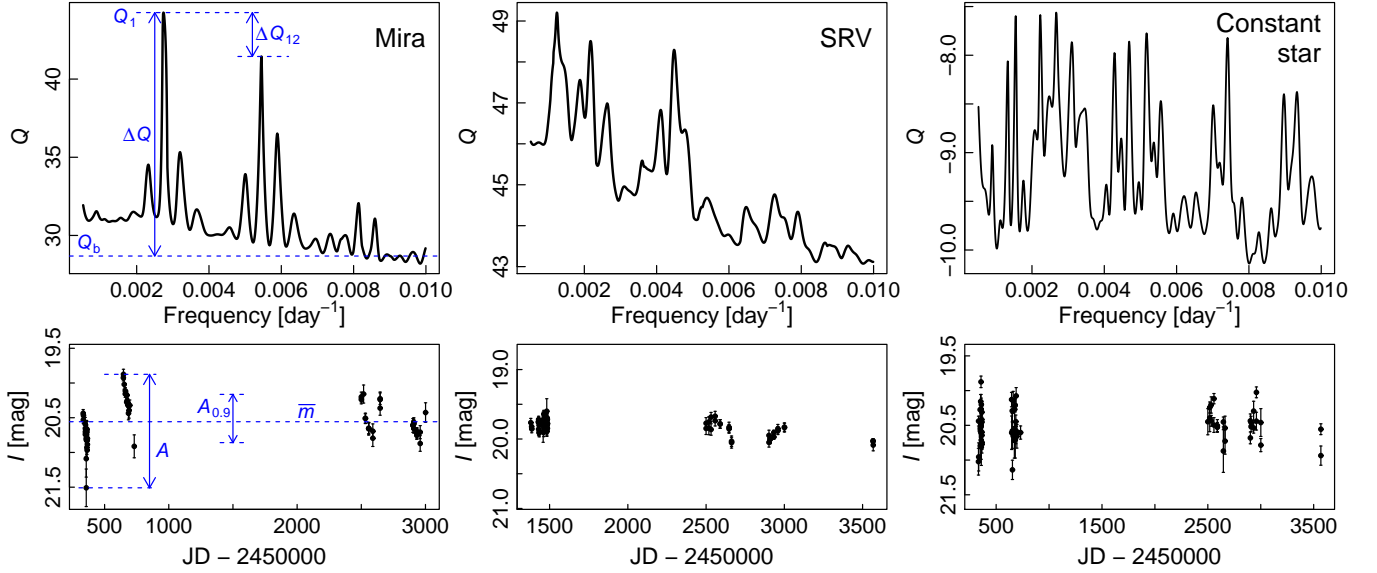}
\caption{Examples of frequency spectra (top) and corresponding light curves (bottom) for a simulated Mira (left), SRV (middle), and ``constant'' star (right). The blue dashed lines and arrows indicate some of the quantities used as classification features.\label{fig:spec}}
\end{figure*}

\begin{deluxetable*}{llcrr}
\tablecaption{Features for the Classifier \label{tab:ftr}}
\tabletypesize{\footnotesize}
\tablewidth{0pt}
\tablehead{
\multicolumn{1}{l}{Feature} & \multicolumn{1}{l}{Description}                                               & \colhead{Src$^a$} & \multicolumn{2}{c}{Rank$^b$}\\
\colhead{}              & \colhead{}                                                                        &                  & \colhead{Mira} & \colhead{C/O}}
\startdata
$\Delta Q$              & Difference of log-likelihoods $Q_1 - Q_b$ (see below) \dotfill                    & F &  1 & 12 \\ [-3pt]
$\sigma(R_q)/\sigma(m)$ & Ratio of standard deviations (see below) \dotfill                                 & L &  2 &  9 \\ [-3pt]
$A_{0.9}$               & Light curve range from 10$^{\rm th}$ to 90$^{\rm th}$ percentile \dotfill     & L &  3 &  5 \\ [-3pt]
$\Delta Q_{12}$         & Difference in log-likelihood between highest and second peak \dotfill  & F &  4 & 11 \\ [-3pt]
$A$                     & Light curve range \dotfill                                                    & L &  5 &  7 \\ [-3pt]
$A_P$                   & Amplitude of the periodic component \dotfill                                      & M &  6 &  2 \\ [-3pt]
$\log{\theta_2}$        & Log of hyperparameter $\theta_2$ \dotfill                                         & M &  7 &  6 \\ [-3pt]
$\sigma(\overline{m})$  & Standard deviation of residuals about $\overline{m}$ \dotfill                     & L &  8 &  4 \\ [-3pt]
$f_1$                   & Best-fit frequency \dotfill                                                       & M &  9 &  1 \\ [-3pt]
$\sigma(R_q)$           & Standard deviation of residuals from piece-wise quadratic fits \dotfill           & M & 10 & 13 \\ [-3pt]
$Q_1$                   & Best-fit log-likelihood \dotfill                                                  & F & 11 & 15 \\ [-3pt]
$\overline{m}$          & Unweighted mean magnitude \dotfill                                                & L & 12 &  3 \\ [-3pt]
$Q_b$                   & The baseline value of frequency spectrum (10$^{\rm th}$ percentile) \dotfill      & F & 13 & 14 \\ [-3pt]
$\sigma(R_{\rm model})$ & Standard deviation of the best-fit model residuals \dotfill                       & L & 14 & 10 \\ [-3pt]
$N$                     & Number of measurements \dotfill                                                   & L & 15 & 18 \\ [-3pt]
$\theta_1$              & Hyperparameter $\theta_1$ \dotfill                                                & M & 16 &  8 \\ [-3pt]
$\sigma(\beta_1)$       & Posterior uncertainty of $\beta_1$ \dotfill                                       & M & 17 & 17 \\ [-3pt]
$\sigma(\beta_2)$       & Posterior uncertainty of $\beta_2$ \dotfill                                       & M & 18 & \mcr{16}
\enddata
\tablecomments{(a): Source of parameter, F=Frequency spectrum, L=Light curve, M=Model. (b): Rank in importance for classification as Mira or discrimination between C- and O-rich subtypes. Rank is determined by the mean decrease in the Gini index.}
\end{deluxetable*}

\noindent{spectrum (hereafter, $f_1$) as the most likely frequency. We found that the true period was successfully recovered (with a tolerance of $|f_1-f_{\rm true}| < 2.7\times 10^{-4}$ as defined in H16) for 69.4\% of all simulated Mira light curves. }

We estimated period uncertainties for all light curves using a non-parametric bootstrap approach followed by error scaling, as follows. First, we resampled the measurements with replacement and derived new values of $f_1$, repeating this procedure 500 times per variable. We used the standard deviation of the results for each object as an initial estimate of the period uncertainty. Next, we carried out the same procedure on the simulated Miras (with 30 iterations per light curve) and calculated $\delta P = (P_i - P_r) / \sigma(P_r)$, where $P_i$ and $P_r$ are the input and recovered periods and $\sigma(P_r)$ is the bootstrap-based uncertainty for the latter. Restricting our analysis to the successfully-recovered variables (as defined in the previous paragraph) and under the assumption that period residuals should follow a Gaussian distribution, we calculated the fraction with $|\delta P|<1$ and iteratively rescaled $\sigma(P_r)$ until 68.3\% of the objects met that criteria. This required a rescaling factor of 2.33, which was then applied to the bootstrap-based uncertainty estimates of the Mira candidates.

\section{Results}\label{sec:results}

\subsection{Random Forest classification of Miras}\label{ssec:rfcls}

Random Forest is a machine-learning technique that has already proven to be effective in classifying different classes of variable stars \citep{Richards2011, Dubath2011}. We built a Random Forest classifier based on the model parameters, the features of the frequency spectra, and information obtained from the simulated light curves as detailed below. Once trained on the simulated data, it was applied to the M33 observations to select Mira candidates. Our choice of Random Forest is supported by a comparative study reported in \S\ref{ssec:rfcomp}, where it is shown to outperform several state-of-the-art classifiers on simulated data. Fig.~\ref{fig:spec} shows the frequency spectra for a representative artificial light curve of each of the three classes (Mira, SRV and ``constant''). The frequency spectra of SRVs and ``constant'' stars are usually quite different due to their lack of periodicity, which indicates the shape of this function can be used to identify Mira candidates.

\begin{figure}[t]
\includegraphics[width=0.48\textwidth]{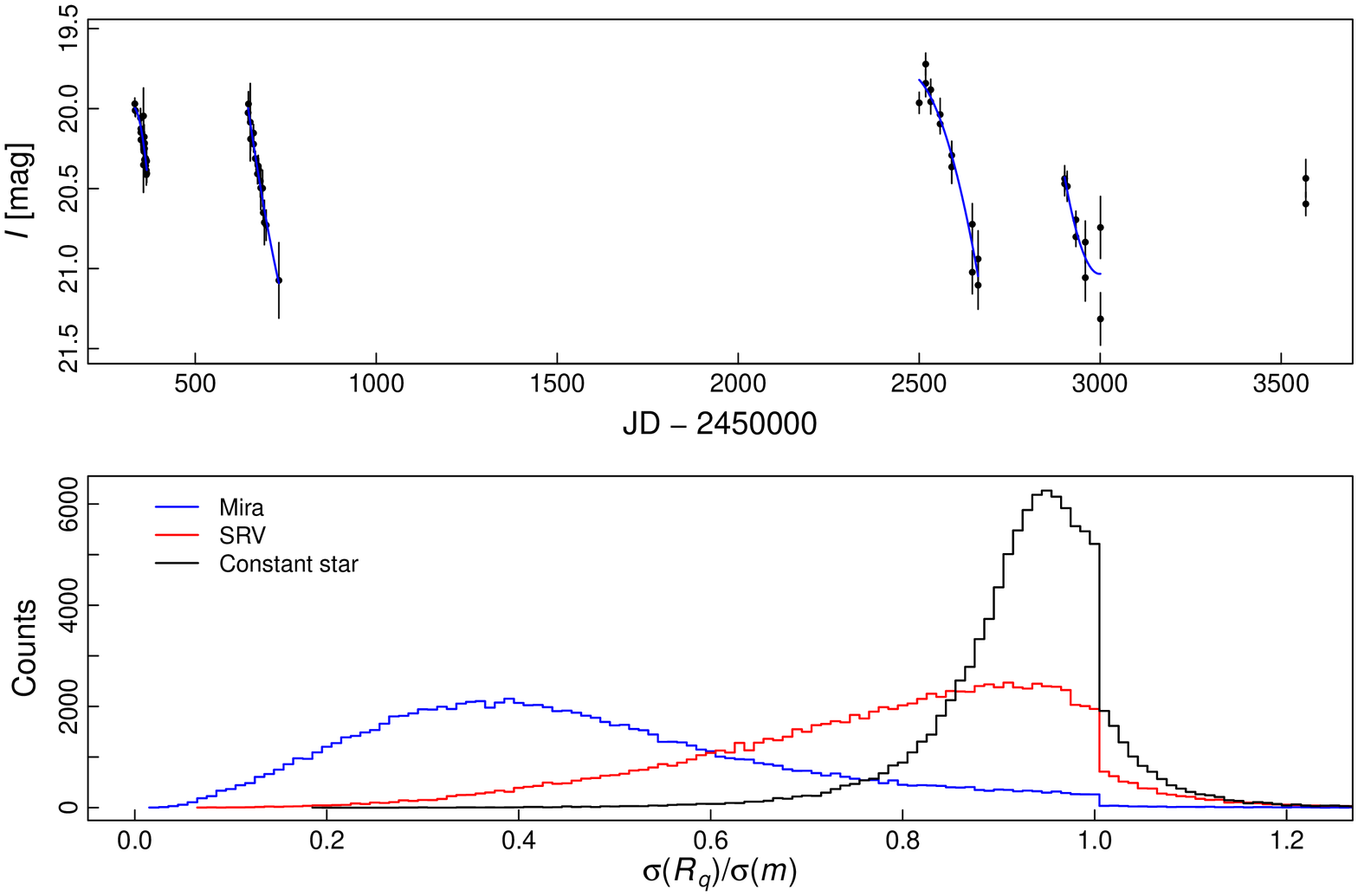}
\caption{{\it Top}: Piecewise quadratic fit to a simulated Mira light curve. {\it Bottom:} Distribution of a classification feature based on such fits for simulated Miras (blue), SRVs (red), and constant stars (black).\\ \\ \label{fig:quand}}
\vspace*{-18pt}
\end{figure}

We extracted 7 features from the best-fit model parameters, 4 from the frequency spectra, and 7 from the light curves. Table~\ref{tab:ftr} provides a summary of all features and their rank in terms of importance for separating Mira candidates from other stars and for separating the

\begin{figure}[t]
\includegraphics[width=0.48\textwidth]{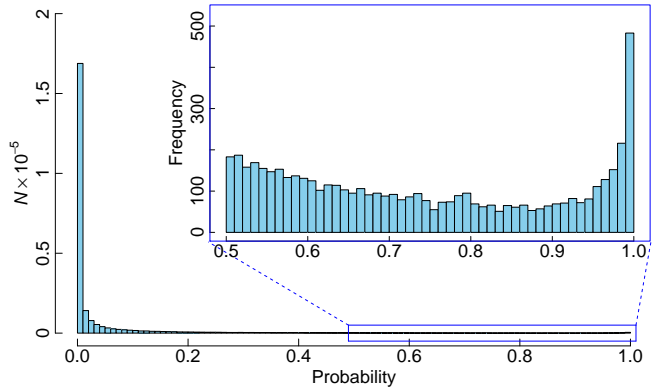}
\caption{Distribution of Random Forest voted values of Mira probability ($P_M$) for the entire M33 sample. There are 5480 objects with $P_M>0.5$. \label{fig:prob}}
\end{figure}

\begin{figure}[h]
\includegraphics[width=0.48\textwidth]{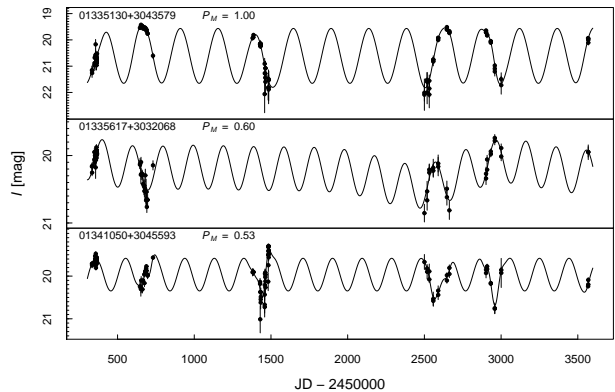}
\caption{Example light curves and best-fit models (solid lines) for likely Miras in M33 with different values of $P_M$. \label{fig:lc}}
\end{figure}

\begin{figure*}[t]
\includegraphics[width=0.33\textwidth]{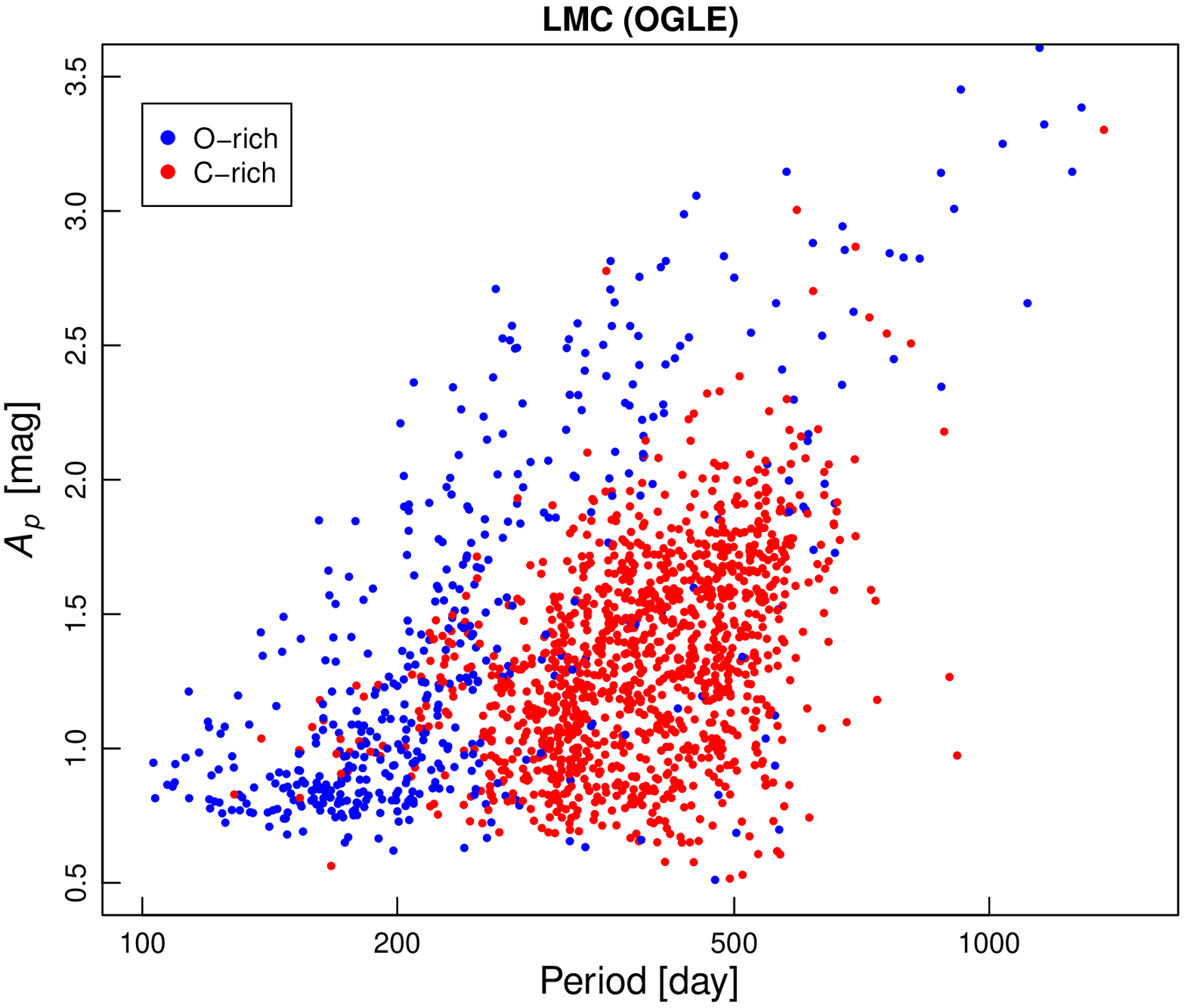}
\includegraphics[width=0.33\textwidth]{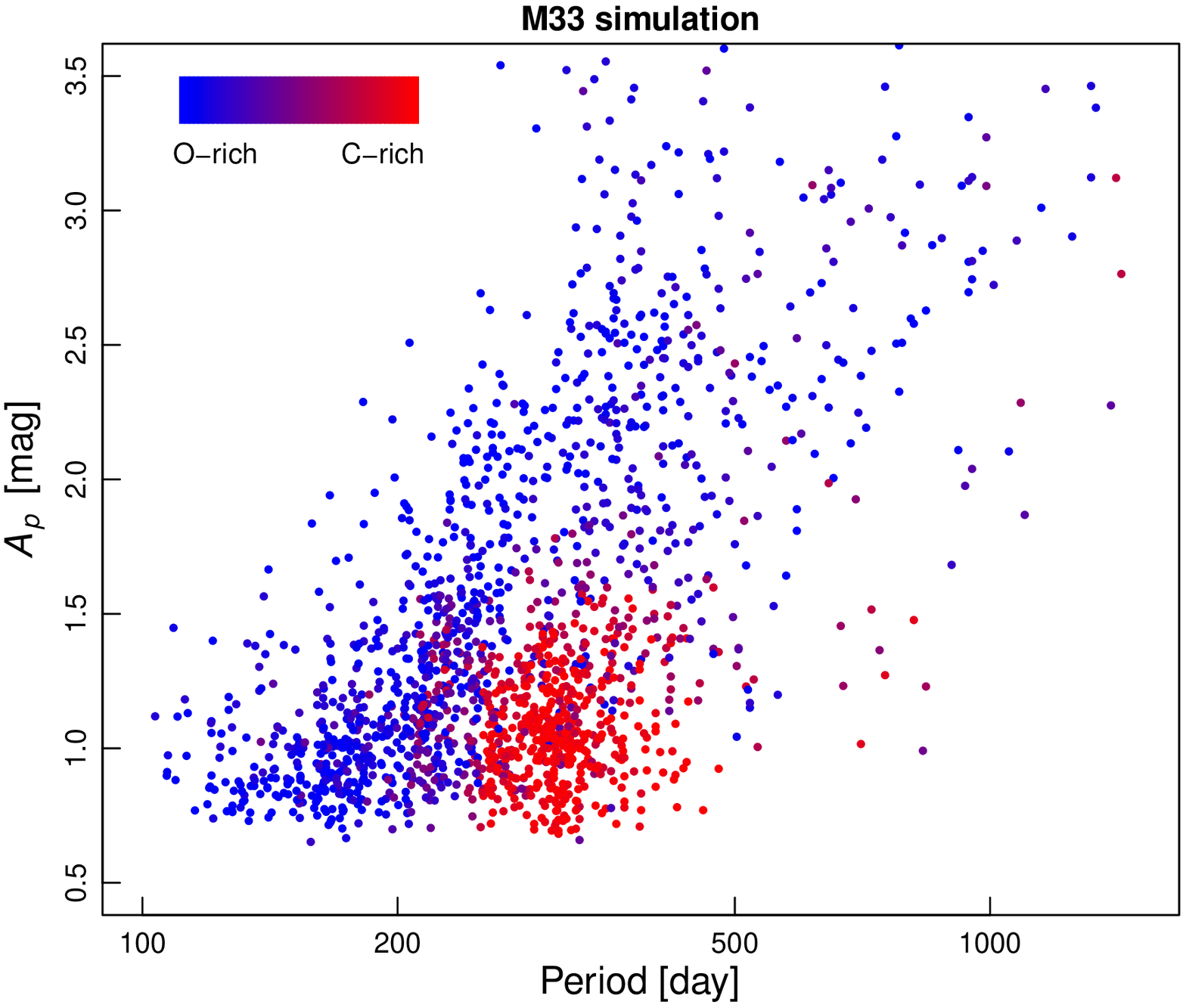}
\includegraphics[width=0.33\textwidth]{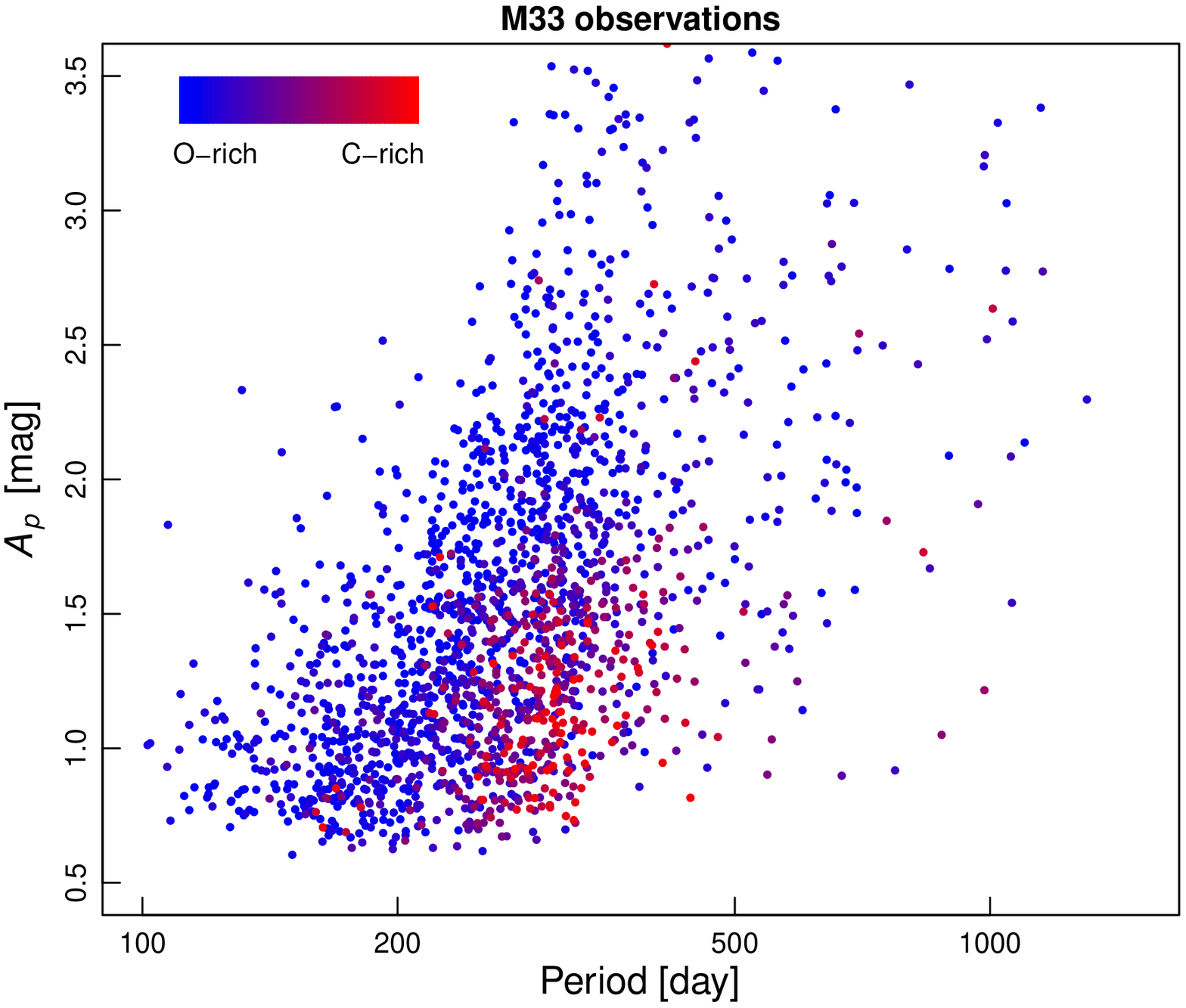}
\caption{Random Forest classification of Mira candidates into O- (blue) or C-rich (red), plotted as a function of $P$ and $A_P$. Left: LMC variables classified by \citet{Soszynski2009}. Middle: simulated M33 variables, based on the LMC sample but accounting for the shallower depth in absolute magnitude of our survey. Right: Mira candidates in M33 from this work.\label{fig:coclass}}
\end{figure*}

\begin{figure*}
\includegraphics[width=\textwidth]{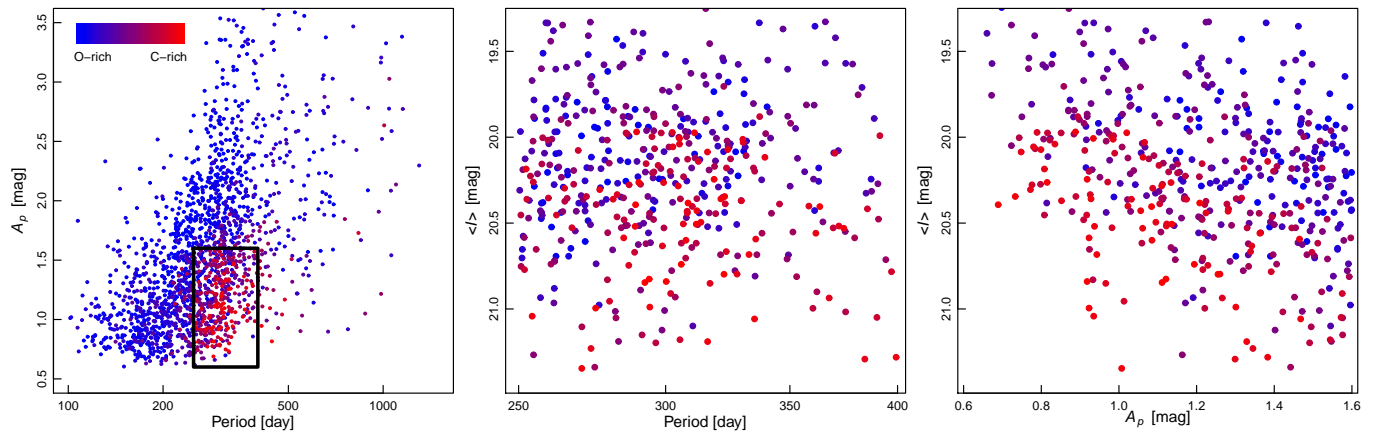}
\caption{Illustration of how multiple attributes help to discriminate O-rich from C-rich Mira candidates. Left: Same as right panel of Fig.~\ref{fig:coclass}, but indicating area of interest where both subtypes overlap. Middle and right: separation of candidates on other two-dimensional slices of the Random Forest parameters.\label{fig:coclass2}}
\end{figure*}

\begin{figure*}
\includegraphics[width=0.495\textwidth]{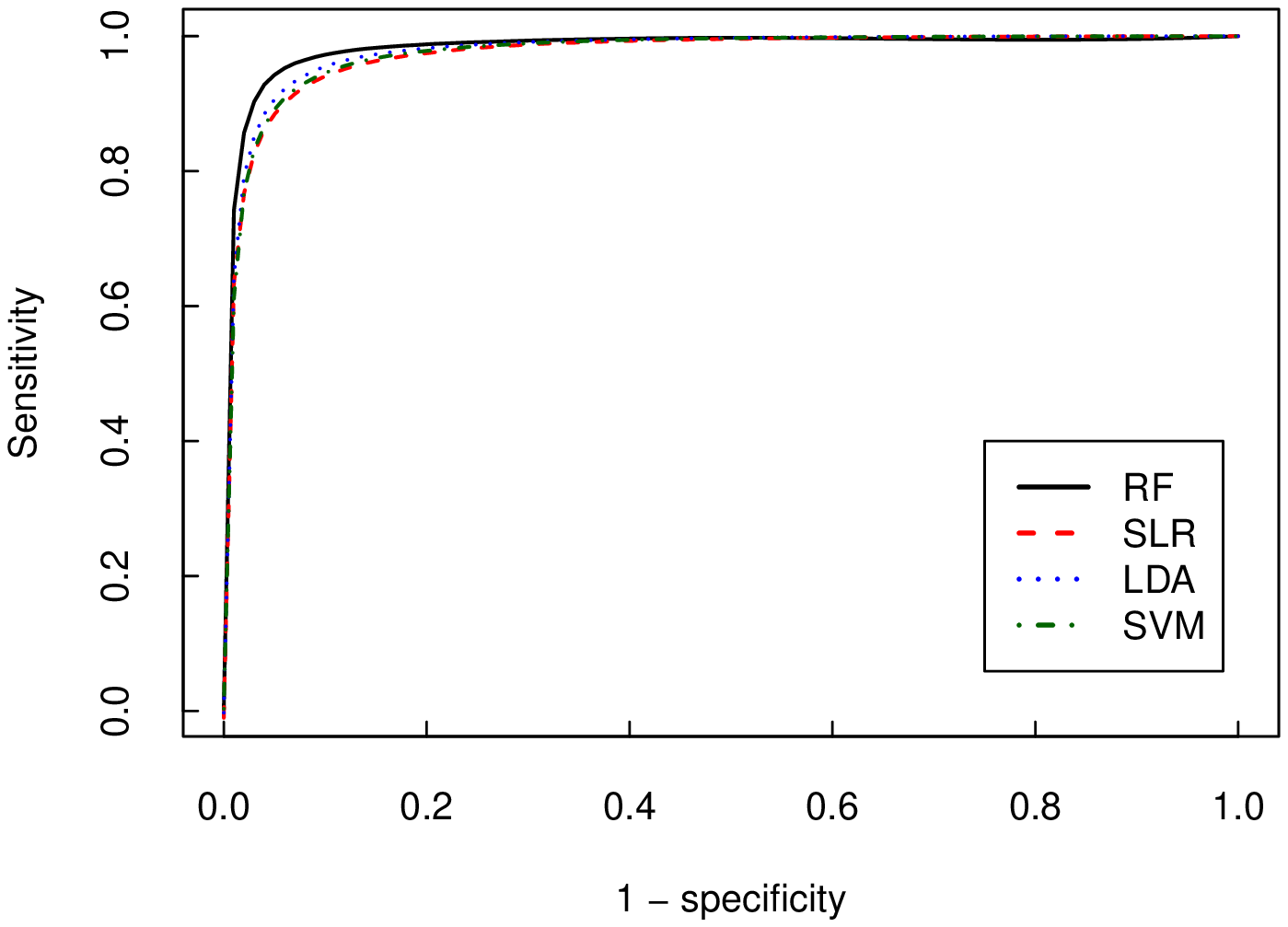}
\includegraphics[width=0.495\textwidth]{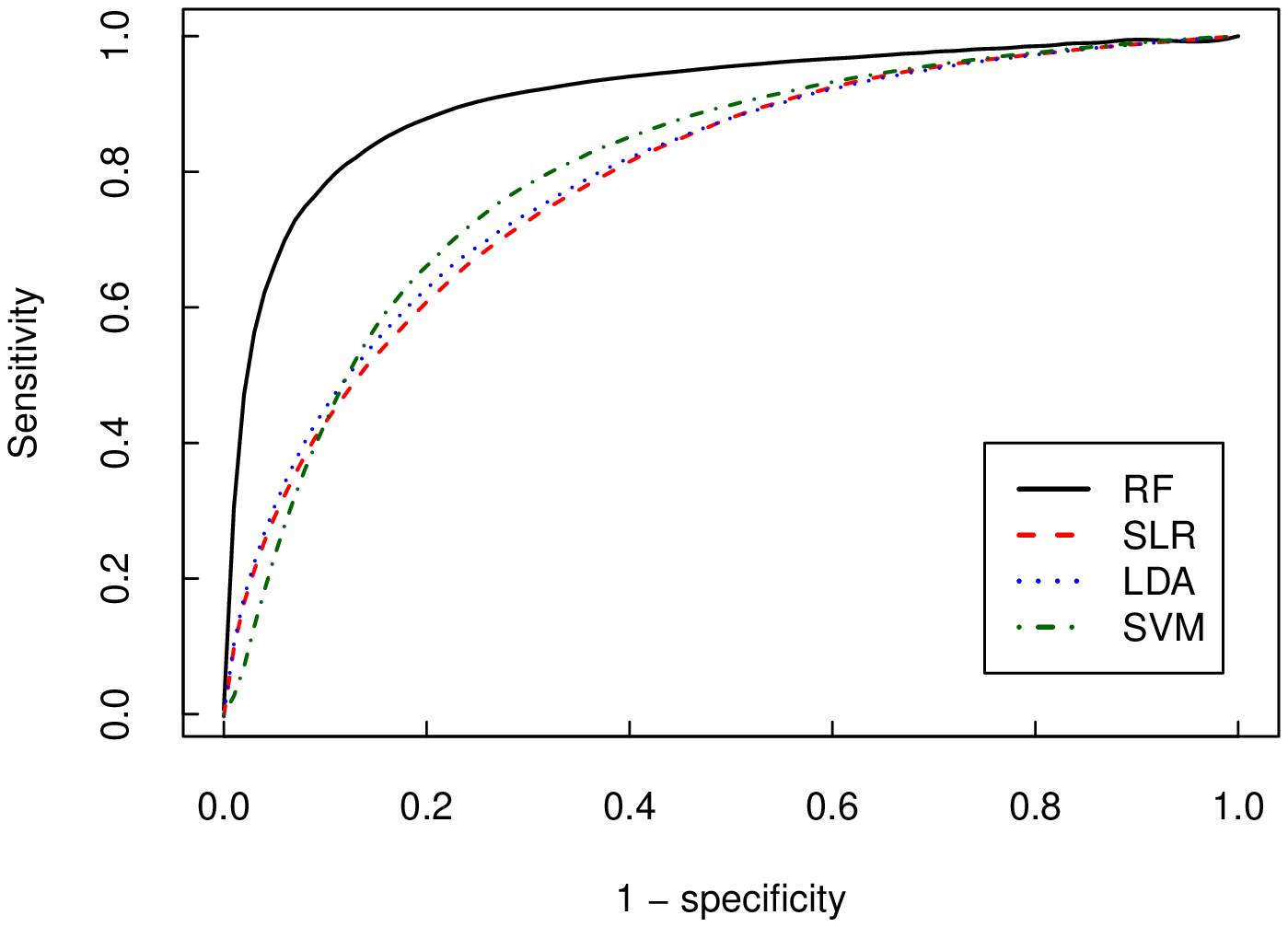}
\caption{ROC curves for classification between Mira/non-Mira (left) and C/O-rich (right) for various classifiers: RF, black solid line; LDA, blue dotted line; SLR, red dashed line; SVM, green dash-dotted line. While all classifiers have very similar AUCs for the first classification, RF significantly outperforms the others in the latter.} \label{fig:rocs}

\end{figure*}

\noindent{former into likely C-rich or likely O-rich. The upper-left panel of Fig.~\ref{fig:spec} shows a graphical representation of the features that were extracted from the frequency spectra. Fig.~\ref{fig:quand} shows two features we extracted from the light curves: the standard deviation of the residuals from piecewise quadratic fits ($\sigma(R_q)$), and its ratio to the scatter about the overall mean value ($\sigma(\overline{m})$). This ratio is significantly smaller for Miras than for any of the other classes.}

We trained the Random Forest classifier by building 400 decision trees with simulated light curves, each composed of $1.2\times 10^5$ non-Miras and $2.5\times 10^3$ Miras. This ratio was chosen to match the estimated fraction of Mira candidates in the actual data, derived from the ratio of Miras to other stars in the OGLE catalog (after applying the M33 completeness function). We then applied the trained classifier to the actual M33 data and obtained the voted probabilities for each star to be a Mira (hereafter, $P_M$); Fig.~\ref{fig:prob} shows the resulting histogram. Based on a five-fold cross validation on the simulated light curves, the Mira recovery rate at $P_M=0.5$ is 75.4\% and the impurity is 0.7\%. There are 5480 objects with probabilities  above this value, 5145 of which have $A_P>0.6$~mag and were selected for further study. Fig.~\ref{fig:lc} shows three representative light curves for Mira candidates with different values of $P_M$. The full set of light curves is available in the online edition of this article. The Mira subtype was tentatively inferred by using another Random Forest classifier trained on the same features, which yielded the probability of each candidate being O-rich ($P_O$). Using the features of Mira candidates in the LMC bar to classify variables in the inner disk of M33 should be a robust approach, given the similar chemical abundances of both regions \citep{Romaniello2008,Bresolin2011}. Fig.~\ref{fig:coclass} shows the separation between subtypes based on the features with the highest rank in terms of discrimination: $P$ and $A_P$. The difference in the distribution of variables between the left and middle panels is due to the shallower depth and sparser sampling of the M33 survey relative to the OGLE coverage of the LMC. We caution that this is a limited two-dimensional view of a classification process that is based on 18 features. Fig.~\ref{fig:coclass2} attempts to provide additional insight into the Random Forest classification process by plotting the distribution of a subsample of candidates in other two-dimensional slices of parameter space. Based on a five-fold cross-validation on the simulated light curves, the O-rich recovery rate at $P_O=0.5$ is 91.4\% and the impurity is 12.8\% while the corresponding values for C-rich variables are 82.3\% and 12.1\%.

\subsection{Comparison with other classification methods}\label{ssec:rfcomp}

Although we chose Random Forest (RF) as our classifier, it is insightful to compare its performance against other popular classifiers. We selected three state-of-the-art classifiers: sparse linear discriminant analysis with $\ell_1$ penalty (LDA), sparse logistic regression with $\ell_1$ penalty (SLR), and a $\nu$-classification support vector machine with radial basis kernel (SVM). We used the same input features discussed in \S\ref{ssec:rfcls}, normalized to zero mean and unit variance.

\begin{deluxetable*}{lrrrrrrrrrrllrrrr}
\tabletypesize{\tiny}
\tablecaption{Mira candidates\label{tab:par}}
\tablewidth{0pt}
\tablehead{
\multicolumn{1}{l}{ID}       & \colhead{R.A.} & \colhead{Dec.} &  \colhead{$P$} & \colhead{$\sigma(P)$} & \colhead{$\langle I \rangle$} & \colhead{$\sigma(\langle I \rangle$} & \colhead{$A_P$} & \colhead{$\sigma(A_P)$} & \colhead{$I_m$} & \colhead{$\sigma(I_m)$} & \colhead{C/O} & \colhead{$N$} & \colhead{$A$} & \colhead{$\sigma(A)$} & \colhead{$\Delta t$} & \colhead{T$_0$} \\
\multicolumn{1}{l}{[M33SSSJ]} & \multicolumn{2}{c}{(J2000) [deg]} & \multicolumn{2}{c}{[d]} & \multicolumn{6}{c}{[mag]} & \colhead{} & \colhead{} & \multicolumn{2}{c}{[mag]} & \multicolumn{2}{c}{[d]}}
\startdata
01321114+3032588 & 23.04642 & 30.54967 & 324.1 &  2.0 & 19.85 & 0.14 & 2.12 & 0.29 & 18.60 & 0.36 & O & 39 & 3.0 & 0.3 & 1955.9 & 2106.1\\
01321450+3019349 & 23.06041 & 30.32637 & 309.7 & 10.0 & 20.05 & 0.07 & 1.70 & 0.09 & 19.55 & 0.08 & O & 46 & 2.2 & 0.2 & 1914.0 & 1993.3\\
01321654+3025260 & 23.06890 & 30.42388 & 295.9 & 11.3 & 20.07 & 0.05 & 0.78 & 0.05 & 19.55 & 0.05 & C & 56 & 1.4 & 0.1 & 1955.8 & \mcr{2074.7}
\enddata
\tablecomments{Table~\ref{tab:par} is published in its entirety in the electronic version of this article. A portion is shown here for guidance on its form and content.}
\end{deluxetable*}

First we considered the classification task of Mira vs. non-Mira. The comparison was in terms of receiver operating characteristic (ROC) and its summary statistic AUC (area under curve). They were computed via repeated splitting of the simulated data set. On each instance, $10^4$ Mira and non-Mira light curves (3.7\% of the total) were sampled without replacement to serve as training data, while the rest served as test data. The procedure was repeated 200 times and the final prediction for each light curve was calculated from the averaged probability across all iterations. The resulting ROC curves, shown in the left panel of Fig.~\ref{fig:rocs}, are nearly identical with AUC values of 0.984, 0.979, 0.975 and 0.976 for RF, LDA, SLR \& SVM, respectively.

\vspace{3pt}

We carried out a similar comparison for the classification task of Miras into C-rich vs.~O-rich, with ROC curves plotted in the right panel of Fig.~\ref{fig:rocs}. In this case RF is significantly superior to the other methods, with AUCs of 0.912, 0.793, 0.787 and 0.801 for RF, LDA, SLR \& SVM, respectively.

\vspace{3pt}

\subsection{Mira candidates and Period-Luminosity Relations}

We examined the distribution of best-fit periods for the selected objects and found a large peak at $P\sim340$~d which is not seen in the LMC samples. Visual examination of the light curves in this period bin showed they exhibit a significant change in the mean magnitude of segments obtained from different telescopes. Further examination of the reference images for each field revealed that for these objects, the poorer angular resolution of the FLWO images resulted in the blending of several sources (clearly separated in the WIYN frames).  We visually inspected each light curve and its respective reference images and rejected affected objects.

\vspace{3pt}

Our final sample consists of 1847 Mira candidates. Table~\ref{tab:par} lists their following properties (and their uncertainties, when applicable): IAU-standard ID,  coordinates,  most likely period, mean $I$~mag, amplitude of the periodic component ($A_P$), brightest magnitude of the best-fit model light curve ($I_m$), subtype (O/C), number of light curve measurements ($N$), range of magnitudes ($A$) and times ($\Delta t$) spanned by the light curve, and time of maximum light for the periodic component ($T_0$). 1581 \& 266 objects were classified as O- \& C-rich, respectively. 

\vspace{3pt}

\begin{figure}[h]
\begin{center}
\includegraphics[width=0.495\textwidth]{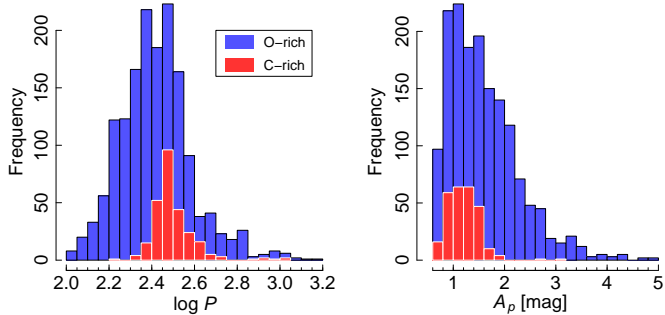}
\end{center}
\caption{Distribution of periods (left) and amplitudes (right) for Mira candidates of each subtype.} \label{fig:peramphist}
\vspace{-12pt}
\end{figure}

\begin{figure}[h]
\begin{center}
\includegraphics[width=0.45\textwidth]{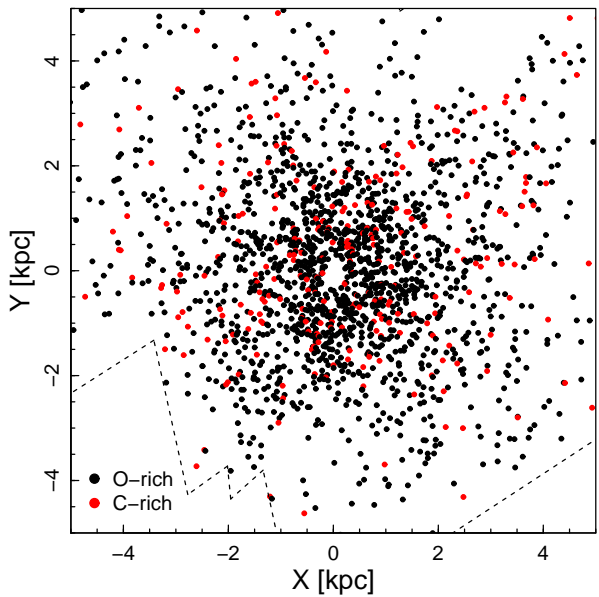}
\end{center}
\caption{Deprojected distribution of Mira candidates (O-rich in black, C-rich in red). The dashed lines indicate the boundaries of our survey.} \label{fig:xydist}
\vspace{-12pt}
\end{figure}

Fig.~\ref{fig:peramphist} shows histograms of periods and amplitudes for both subtypes, while Fig.~\ref{fig:xydist} shows their deprojected galactocentric distribution. Our survey is limited to the innermost $\sim 5$~kpc of the galaxy and within this limited area we see no statistically significant difference in the distribution of candidates by subtype or period. 

\vfill\pagebreak\newpage

We note that we recovered the only spectroscopically-confirmed Mira in M33 \citep{Barsukova2011}, for which we found $P=578\pm32$~d (in contrast to the previously published estimate of $P=665$d). Our classifiers yielded $P_M=0.89$ and $P_O=0.9$ for this object.

\begin{figure*}[htbp!]
\begin{center}
\includegraphics[width=0.95\textwidth]{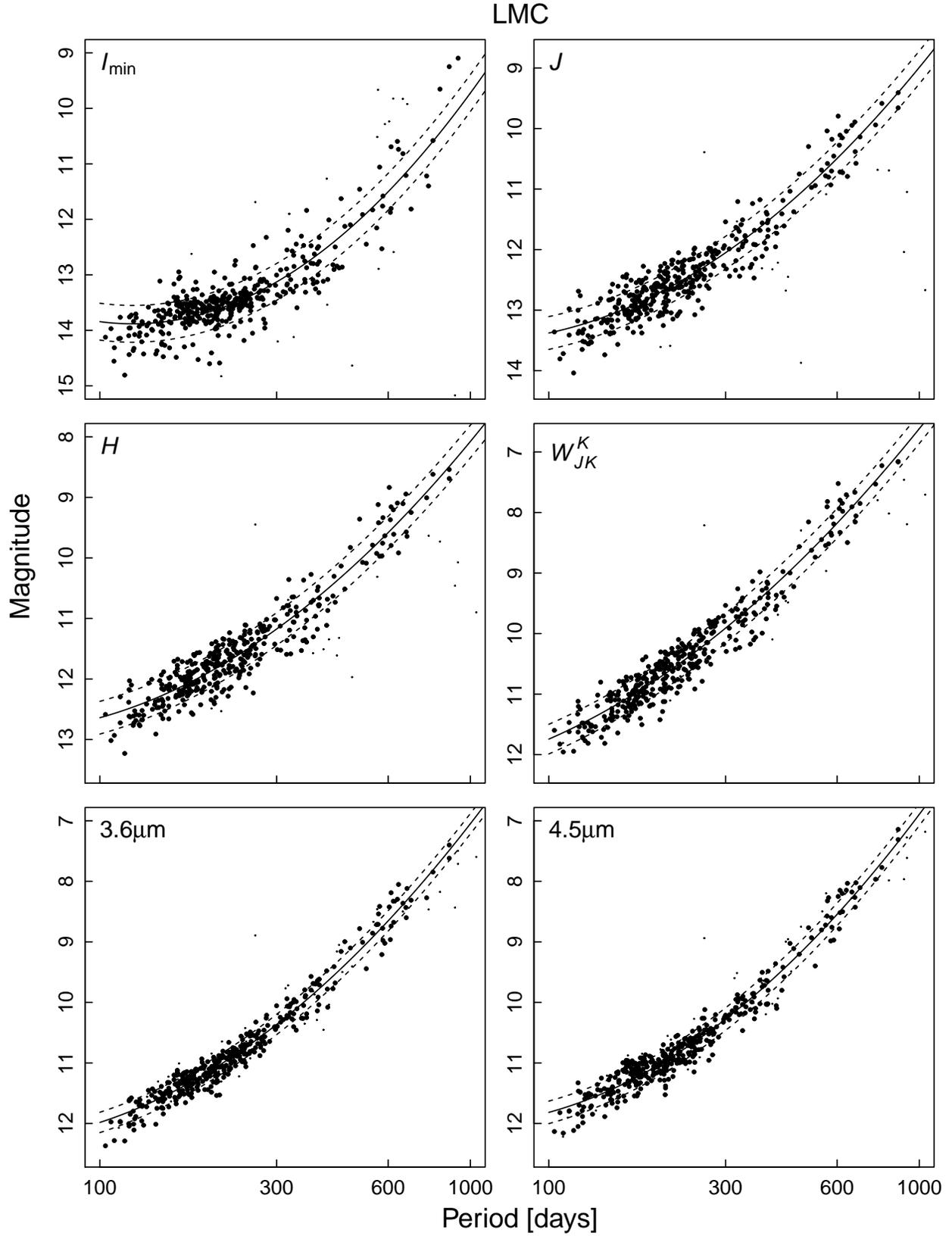}
\end{center}
\caption{PLRs in several bands for Mira candidates classified as O-rich in the LMC. The solid lines show the best-fit quadratic relations to the final LMC samples (large symbols) after iterative $3\sigma$ clipping of outliers (small dots). Dashed lines indicate the dispersion in the fits. \label{fig:pllmc}}
\end{figure*}

\begin{figure*}[htbp!]
\begin{center}
\includegraphics[width=0.95\textwidth]{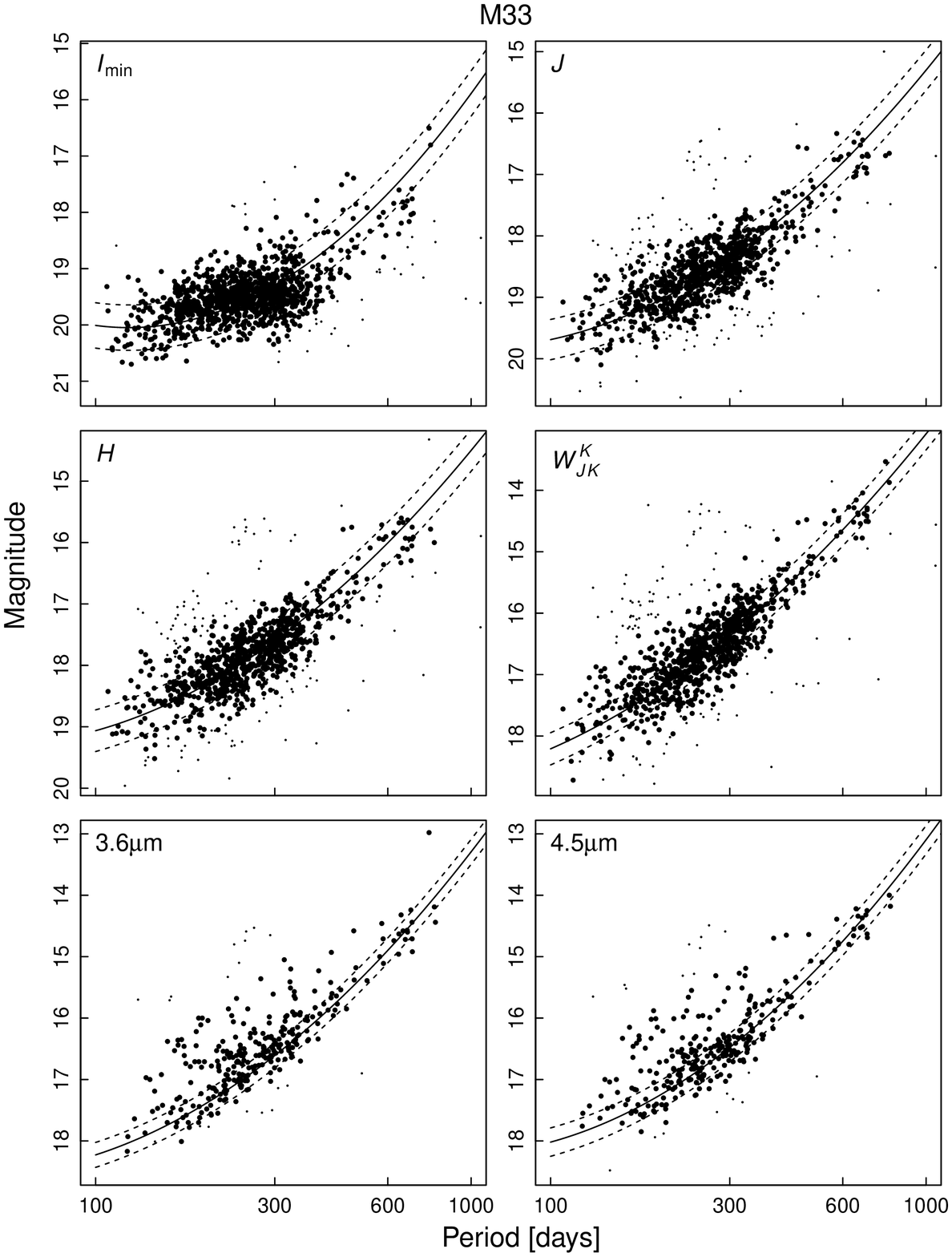}
\end{center}
\caption{PLRs in several bands for Mira candidates classified as O-rich in M33. The solid lines show the LMC-based quadratic relations of Fig.~\ref{fig:pllmc}, shifted by the best-fit relative distance modulus in each band (including blending correction). Small dots indicate variables removed by iterative $3\sigma$ clipping. Dashed lines indicate the dispersion of the Gaussian component of the model.\label{fig:plm33}}
\end{figure*}

Figs.~\ref{fig:pllmc} \& \ref{fig:plm33} show preliminary PLRs for O-rich Mira candidates in the LMC and M33 at wavelengths ranging from 0.8 to 4.5$\mu$m. We emphasize the following is a simple analysis to demonstrate the validity of our methods for identifying, phasing and classifying Mira candidates in M33. A complete analysis (including C-rich candidates) will be presented in a future paper.

The $I_m$ magnitudes of the LMC Mira candidates were determined using the method described in \S\ref{sec:simulation}, while the random-phase magnitudes at longer wavelengths were obtained from the SAGE catalog \citep{Meixner2006}. We chose to plot the $I_m$ PLR to show a minimally biased comparison of the relations at this wavelength, since the $V\!-\!I$ colors necessary to generate a ``Wesenheit''-corrected mean-light $I$-band PLR are not available for M33. We show one example of a relation corrected for interstellar extinction for $K_s$, using the formulation of \citet{Soszynski2009}. We note that this formulation may not be appropriate to correct for the circumstellar dust that is specially prevalent among C-rich and long-period Miras \citep[see][for a thorough analysis of this issue]{Ita2011}. We solved for quadratic PLRs,
\begin{equation}
m = a_0 + a_1 (\log P - 2.3) + a_2 (\log P - 2.3)^2
\end{equation}
\noindent{using an iterative $3\sigma$ clipping and removing the single largest outlier in each band until convergence. Table~\ref{tab:plr} summarizes the results of the fits.}

The M33 sample was restricted to 1161 candidate variables with $A_P/A < 1.1$, $\sigma(A_P)/A_P < 0.15$, $\sigma(P)/P<0.1$ and $P<\Delta t$. These selection criteria were based on an examination of the input and recovered parameters for the simulated M33 Miras. When the amplitude of the periodic component significantly exceeds the range of magnitudes spanned by the data, and/or the best-fit period is longer than the time span of the light curve, the recovered parameters exhibit considerably larger scatter and the fraction of variables with successfully recovered periods (as defined in H16) is noticeably lower. The simulated O-rich Miras that met our selection criteria had input/output ratios of $A_P$ of $1.02\pm 0.21$, versus $0.70\pm 0.82$ for the others. Likewise, the fraction of successfully recovered periods was 86\% for the variables meeting the criteria versus only 45\% for the others. 

The $I_m$ magnitudes of M33 Mira candidates are from Table~\ref{tab:par}, while the random-phase magnitudes at longer wavelengths were taken from \citet[for $JHK_s$]{Javadi2015} and \citet[for 3.6 \& 4.5$\mu$m]{Thompson2009}. We matched the catalogs using tolerances of $0\farcs 3$ and $0\farcs 5$ and found 972 and 302 counterparts, respectively. We fixed the linear and quadratic terms of the PLRs to those derived from the LMC sample and solved for $a_0$, applying an iterative $3\sigma$ clipping that removed the single largest outlier at a time. Once this procedure converged, we modeled the cumulative distribution of PLR resid uals as the combination of a Gaussian (to account for the finite width of the instability strip) plus an exponential distribution towards brighter values (to account for blends, which can only bias the residuals in one direction). The final values of $a_0$ listed in Table~\ref{tab:plr} include these ``blending corrections'', which amount to $\sim0.09$~mag and $\sim0.25$~mag at near- and mid-infrared wavelengths, respectively. The larger contamination for the two longest bands is likely due to the significantly poorer\ \ angular resolution\ \ of the {\it Spitzer} images.\ \ The

\begin{deluxetable*}{llrrrrrr}
\tabletypesize{\small}
\tablecaption{Preliminary O-rich Mira Period-Luminosity relations\label{tab:plr}}
\tablewidth{0pt}
\tablehead{
\colhead{Galaxy} & \colhead{$\lambda$} & \colhead{$a_0$} & \colhead{$a_1$} & \colhead{$a_2$} & \colhead{$\sigma$} & \colhead{$N_i$} & \colhead{$N_f$}}
\startdata
\multirow{7}{*}{LMC} & $I_m$ & 13.66$\pm$0.02 & -2.12$\pm$0.14 & -5.01$\pm$0.37 & 0.33 &  427 &  399 \\\cline{7-8}
                     & $J$   & 12.71$\pm$0.01 & -3.16$\pm$0.01 & -3.07$\pm$0.03 & 0.27 & \multirow{6}{*}{390} & \multirow{6}{*}{370} \\
                     & $H$   & 11.87$\pm$0.01 & -3.42$\pm$0.01 & -2.85$\pm$0.03 & 0.27 & & \\
                     & $K_s$ & 11.52$\pm$0.01 & -3.72$\pm$0.01 & -2.75$\pm$0.03 & 0.24 & & \\
                     & $W^K_{JK}$ & 10.72$\pm$0.01 & -4.15$\pm$0.02 & -2.46$\pm$0.04 & 0.25 & & \\
                     & 3.6   & 11.12$\pm$0.01 & -3.75$\pm$0.01 & -2.95$\pm$0.03 & 0.17 & & \\
                     & 4.5   & 11.02$\pm$0.01 & -3.63$\pm$0.01 & -3.24$\pm$0.03 & 0.19 & & \\\cline{1-8}
\multirow{7}{*}{M33} & $I_m$ & 19.82$\pm$0.01 &        \nd       &        \nd       & 0.40 & 1161 & 1125 \\\cline{7-8}
                     & $J$   & 19.02$\pm$0.01 &        \nd       &        \nd       & 0.33 & \multirow{4}{*}{972} & \multirow{4}{*}{853} \\
                     & $H$   & 18.29$\pm$0.01 &        \nd       &        \nd       & 0.34 & & \\
                     & $K_s$ & 17.94$\pm$0.01 &        \nd       &        \nd       & 0.26 & & \\
                     & $W^K_{JK}$ & 17.19$\pm$0.01 &        \nd       &        \nd       & 0.32 & & \\\cline{7-8}
                     & 3.6   & 17.20$\pm$0.01 &        \nd       &        \nd       & 0.20 & \multirow{2}{*}{302} & \multirow{2}{*}{282} \\
                     & 4.5   & 17.15$\pm$0.01 &        \nd       &        \nd       & 0.23 & & 
\enddata
\tablecomments{$N_i$: initial number of variables in the sample. $N_f$: final number after iterative outlier rejection. $\sigma$: Gaussian width. The fits to the M33 samples include a blending correction and used fixed values of $a_1$ and $a_2$ determined from the LMC samples.}
\end{deluxetable*}

\noindent{scatter in the M33 PLRs (after accounting for blended objects) compares favorably with the higher-quality LMC samples and the mean (error-weighted) LMC-relative distance modulus of $6.31\pm0.11$~mag is consistent with previous determinations \citep{bonanos06,Pellerin2011}.}

\section{Summary}

We carried out a search for Mira variables in M33 using sparsely-sampled $I$-band light curves. We determined periods using a novel semi-parametric Gaussian Process model and used the Random Forest method to identify Mira candidates and classify them into Carbon- or Oxygen-rich subtypes. We identified 1847 likely Mira candidates, most of them O-rich, which exhibit Period-Luminosity Relations with dispersions comparable to those seen in the Large Magellanic Cloud.

WY \& LMM acknowledge financial support from NSF grant AST-1211603 and from the Mitchell Institute for Fundamental Physics and Astronomy at Texas A\&M University. S.H. was partially supported by the Texas A\&M University-NSFC Joint Research Program. JZH was partially supported by NSF grant DMS-1208952. We thank Drs.~A.~Javadi \& J.~L.~Prieto for providing the M33 photometry at near- and mid-infrared wavelengths, respectively. We thank the anonymous referee for helpful and constructive comments.

This publication has made use of the following resources:

\begin{itemize}

\item data products from the Optical Gravitational Lensing Experiment, conducted by the Astronomical Institute of the University of Warsaw at Las Campanas Observatory, operated by the Carnegie Institution for Science.

\item the VizieR catalogue access tool and the cross-match service provided by the Centre de Donn\'ees astronomiques de Strasbourg, France.

\item NASA's Astrophysics Data System at the Harvard-Smithsonian Center for Astrophysics.

\item the NASA/IPAC Extragalactic Database (NED) which is operated by the Jet Propulsion Laboratory, California Institute of Technology, under contract with NASA.

\item data products from the Two Micron All Sky Survey, which is a joint project of the University of Massachusetts and the Infrared Processing and Analysis Center at the California Institute of Technology, funded by NASA and the NSF.

\item the Texas A\&M University Brazos HPC cluster.
\end{itemize}

\ \par

\facility{FLWO:1.2m,WIYN}
\software{DAOPHOT, ALLSTAR, ALLFRAME, TRIAL \citep{Stetson1987,Stetson1994,Stetson1996}, WCSTools \citep{Mink1999}}

\bibliographystyle{aasjournal}
\bibliography{mira}

\end{document}